# Magnetization reversal assisted by half antivortex states in nanostructured circular cobalt disks

A. Lara[1], O. V. Dobrovolskiy[2,3], J. L. Prieto[4], M. Huth[2], and F. G. Aliev[1*]

[1] Dpto. Física de la Materia Condensada, Universidad Autónoma de Madrid, Madrid, Spain

[2] Physikalisches Institut, Goethe University, Frankfurt am Main, Germany

[3] Physics Department, V. Karazin National University, Kharkiv, Ukraine

[4] Instituto de Sistemas Optoelectrónicos y Microtecnología (ISOM), Universidad Politecnica de Madrid, Spain

The half antivortex, a fundamental topological structure which determines magnetization reversal of submicron magnetic devices with domain walls, has been suggested also to play a crucial role in spin torque induced vortex core reversal in circular disks. Here we report on magnetization reversal in circular disks with nanoholes through consecutive metastable states with half antivortices. In-plane anisotropic magnetoresistance and broadband susceptibility measurements accompanied by micromagnetic simulations reveal that cobalt (Co) disks with two and three linearly arranged nanoholes directed at 45º and 135º degrees (º) with respect to the external magnetic field show reproducible step-like changes in the anisotropic magnetoresistance and magnetic permeability due to transitions between different intermediate states mediated by vortices and half antivortices confined to the dot nanoholes and edges, respectively. Our findings are relevant for the development of multi-hole based spintronic and magnetic memory devices.

(*) Corresponding author: farkhad.aliev@uam.es

Magnetization reversal in small magnetic elements is assisted by topological structures (TSs). A detailed theoretical analysis showed that the TSs usually found in simple magnetic elements can be decomposed into only two topologically different but simple objects: the vortex (V) and the half-antivortex (HAV) [1]. Controlling vortex, antivortex (AV), edge HAV and domain walls (DWs), which in turn can be decomposed into two HAVs with different or equal signs [2], becomes important for both fundamental and technological reasons [2,3]. Thus, recent studies revealed the importance of a controlled introduction of HAVs for effective domain wall displacement, manipulation [2] and dynamic response [4]. It has been found [5] that a DW moving in a finite strip at sufficiently high velocities is subject to the appearance of vortices besides the two half antivortices defining the original DW.

Circularly and elliptically shaped ferromagnetic disks are basic elements of spin torque oscillators [6], magnonic devices [7] and magnetic memories [8-9]. Under certain conditions these circular disks exhibit a single vortex (SV) ground state [8]. A similar state is also created in the free layer of a circular spin-valve structure by a strong spin torque caused by the adjacent perpendicularly magnetized layer [10]. On the one hand, vortex-based hybrid devices may represent a breakthrough for the implementation of spintronics in future telecommunication [11]. On the other hand, metastable vortex states emerging under the action of large spin-polarized currents have been predicted to damp spin torque oscillations [10]. Effectively controlling metastable states in magnetic disks and in particular stabilizing V and AV states [12] within the same device represent important emerging problems in nanomagnetism.

Previously, the structuring of circular disks (mainly drilling holes or introducing notches) has been used for controlling the SV nucleation and propagation [13-20], thus modifying the azimuthal spin wave dynamics [21] or deterministically changing the sense of the vortex chirality [22]. For instance, suitable nanostructuring of permalloy (Py) disks introduces a bi-stable SV switching which was suggested as a basic mechanism for reprogrammable logic and data storage devices [23]. Micromagnetic simulations, however, indicate the important role played by HAVs for increasing the effectiveness of current-induced V switching in spin-torque devices [3]. Previous disks nanostructuring aimed at guiding SV trajectories only (see [4] for example on *guided magnetization reversal through a controlled sequence of ground states*). Different approaches are needed for an effective introduction of nontrivial TSs such as HAVs in circular magnetic disks.

In our work we study the introduction of artificial pinning structures to guide *magnetization reversal in circular disks through metastable magnetic configurations*, which are close to a double vortex state (DV) formed by two half antivortices and two vortices connected via DWs [24,25]. The DV state (proposed as a bit-cell for magnetic solid state data storage [26]), was previously stabilized by enhanced roughness in circular Py disks [24] and by the introduction of notches in elliptical Py disks [26]. Here we introduce hole-nanostructured circular magnetic disks (2000 nm diameter, 30 nm thickness, made of cobalt) as elements where HAVs can be stabilized for a broad range of magnetic fields. By using a set of two or three small centrally symmetric nanoholes directed at 45º and 135º with respect to the external magnetic field we demonstrate a robust magnetization reversal through intermediate states containing HAV and V states.

The samples were prepared on top of silicon (Si) substrates covered with a 2 nm-thick silicon dioxide (SiO$_2$) layer. The substrate was pre-patterned with a layout of 300 nm thick gold contacts as shown in Fig. 1(a). Next, the substrate was mounted into a in a high-resolution dual-beam scanning electron microscope (SEM, FEI Nova Nanolab 600). At the position where the Co disks were later on fabricated, the gold contacts were thinned by focused ion beam (FIB) milling to a remaining gold layer thickness of about 30 nm. This ensured good electrical contact between the electrodes and the Co dot. For the FIB milling process the beam parameters were 30 kV / 10 pA, the dwell time was 1µs, and the pitch was 5 nm. Following the milling step, the Co disks were fabricated by focused electron beam induced deposition (FEBID) employing the precursor Co$_2$(CO)$_8$. Finally, the required number of cylindrical nanoholes of about 70 nm diameter was milled into the Co disks by FIB. SEM images of the samples thus fabricated are shown in Fig. 1(a).

Both anisotropic magnetoresistance (AMR) and dynamic susceptibility measurements were done by contacting the gold terminals to a high frequency probe. For AMR measurements, a dc current was applied, and the voltage drop across the dot was measured with a nanovoltmeter. For dynamic susceptibility measurements, a vector network analyzer (VNA) was used for measuring the reflection parameter $S_{11}$, which is the ratio of the reflected and applied power. This parameter needs to be normalized to a reference value, to remove the influence of cables, gold contacts, etc. Here, we chose as a reference the value of $S_{11}$ at the highest field applied. For further details on broadband measurements see [24,28].

To simulate the AMR curves, the angle $\theta(x,y)$ between the local current and the local magnetization in each simulation cell of a grid structure is needed, as changes in the resistance are proportional to $cos^2\theta$. Static micromagnetic simulations have been done using OOMMF [29] to quantify the magnetization for different applied fields. The current distribution and the corresponding electric field were obtained by numerically solving the 2D Laplace's equation for the electric potential (see supplementary materials [30]). The increase of the resistance was calculated by averaging the local increase of the resistance in each simulation cell.

Figure 1(b) shows exemplarily the simulated time evolution of the total energy of an initially saturated hole-free magnetic dot after reducing the saturating external magnetic field to zero. Before reaching the SV ground state, the dot undergoes a series of metastable states that appear as plateaus in the total energy for a few nanoseconds. In the long-time limit two possible states are observed: a metastable DV and the SV ground state, depending mainly on the value of the Gilbert damping constant α. In our hole nano-structuring we took into account previous studies which found [25] that the line connecting two vortices in the DV state is oriented at approximately 45° with respect to the direction of the saturating field before it was switched off, although larger disks (as in the present case) allow for more possibilities. Based on this, we focused on placing holes at 45° and 135° with respect to the direction of the applied field, with the idea of locating the holes at places where vortices are more likely to move, therefore facilitating their trapping.

Figure 1(b) shows that a number of metastable states with different internal TSs appear in the dot before it finally relaxes to the single SV state. Note that the winding number is conserved, so that antivortices (with value -1) appear alongside vortices (with value +1), to keep this quantity equal to +1, the value corresponding to the SV state. In the transition from states with multiple vortices to the SV state, both Vs and AVs can approach each other and mutually annihilate inside the dot or at the edge. Previous works [13,14] dealt with trapping SV as it moves in the direction perpendicular to the applied field. However, DVs [25] do not usually move along this line, but rather rotate around the center of the dot in more complicated ways. Placing nanoholes off the direction perpendicular to the applied field therefore represents a good approach towards trapping both V and HAV states.

Figure 2(a) shows AMR data of Co disks with three different nanohole structures for which most reproducible results were obtained. The step-like changes suggest that intermediate states are involved with multiple V and HAV states. To verify this assumption, we have done micromagnetic simulations of the magnetic hysteresis of Co disks with nanoholes and complemented these results with AMR calculations shown in Fig. 2(b). Both the two and three nanohole structures show clear transitions between multiple intermediate TSs, as indicated by arrows. The corresponding magnetic states are shown in Fig. 3.

Complementary broadband ac response measurements, also shown in Fig. 2(a) by blue lines, further corroborate the presence of a number of intermediate well-defined magnetic states, correlating reasonably well with the signatures found in the dc measurements. In Fig. 2(a) the ac response consists of an average of the normalized $S_{11}$ parameter over the frequency range covered by the network analyzer (10 MHz to 9 GHz) as a function of the applied field. The data acquired by using both techniques point to the presence of the same intermediate states in the investigated samples.

Micromagnetic simulations allow us to identify the intermediate states which develop in the disks with two or three centrally symmetric nanoholes, see Fig. 3. We have found that these states have multivortex character, containing both Vs and HAVs. They are therefore *qualitatively different* *from those* reported for disks with vortex core removal by a single central nanohole [31] or with nanoholes *designed to capture only a SV* [13].

A comparison of simulations and experimental results shows qualitatively similar jumps in magnetoresistance, occurring at quantitatively different values of applied field. We attribute these differences mainly to the far-from-perfect purity of Co that can be achieved with FIB induced deposition (about 90%) [29]. The unavoidable presence of molecules of the precursor gas inside the disk can represent an extra source of pinning for the magnetic topological structures, being higher fields necessary to release them than if the impurities were not present (as happens in simulations). Additionally, there is an uncertainty in choosing the parameters of the simulation (see description in [30]), that could be relevant for studying both the ground and metastable states. This is shown in Fig 1(b), that exposes the influence of damping on how fast the ground state is reached in simulations. The above discussed factors affecting the trapping of metastable states do not allow for a quantitative reproduction by simulation of the measured characteristic fields that release trapped metastable states. Our simulations, therefore, are mainly useful for identifying the processes involving the interaction of holes and magnetic topological defects, and how this is reflected in AMR. The observed stabilization of multi-vortex intermediate/metastable states indicates that the nanohole arrangement leads to a reduction of energy of DV states via adequate pinning of both vortex cores and DWs by holes. Figure 3 shows representative magnetic configurations for which DWs remain pinned to a nanohole even at the cost of increasing the DW energy by increasing its length. Varying the dot diameter allows for tuning the pinning force experienced by the topological structures, but can also result in different magnetic states during the reversal process [31]. One reason for this is the distortion that nanoholes induce in the magnetization around them, which tends to align the magnetic moments parallel to the nanoholes' edges, just like it happens near any edge, in order to reduce the stray field outside the dot.

Among different possible multivortex states, the DV state [24] composed of two half antivortices and two vortices connected via DWs is the simplest one. Our dynamic simulations [31] show that whenever an unstable DW meets a nanohole, it gets pinned due to the associated decrease in the exchange energy of the part corresponding to the nanohole diameter. Since DWs occupy more space than a vortex core, they are more likely trapped by a nanohole than the vortex core itself. Therefore, due to the change of the local magnetization orientation, the disks with intermediate multivortex states will exhibit abrupt changes of AMR as the field changes due to sudden DW depinning and less likely due to V pinning at nanoholes, as shown in Figs. 2 and 3. Here we state as the main result of this work that the nanohole/field configuration used in our experiments provides a robust pathway towards introducing HAV states in circular disks. The reproducibility of the measurements is quite high; by measuring 20 curves we could check that the transitions are always present.

The proposed method of controlling metastable states in circular magnetic disks is capable of providing a much richer variety of intermediate states than the one used to pin only a SV [13,14]. Since DWs extend all over the dot if edge HAVs are present, several different more complex nanohole configurations are suggested. For example, with a 6-hole configuration forming a hexagon-type ring, our simulations and measurements show that a simpler AMR curve results, with smeared out transitions. The 6-hole configuration traps metastable multivortex states, but releases only a SV at sufficiently high fields, causing the AMR curves to look more similar to those of non-patterned disks, but with much higher remanence [31].

The multihole circular disks considered here could serve as basic elements for multistate memories, similarly to the one suggested previously for elliptic dots with notches [26]. In order to take full advantage of the multilevel resistive states, one could incorporate nanostructured dots with multivortex states as one of the electrodes in magnetic tunnel junctions (MTJs). Controllable DW displacement by perpendicular spin torque [32,33] could stimulate further development of these MTJs-based multilevel magnetic memories. Also, further studies could try not only to control the magnetic properties of devices by placing defects at different positions, but also to optimize the hole size for trapping magnetic structures. The hole should remain as small as possible and still possess an effective trapping capability, so that it weakly affects other device properties, in particular due to the unavoidable tendency of magnetization to align parallel to the hole edge.

In *conclusion*, magnetization reversal in circular cobalt disks with two or three centrally symmetric nanoholes takes place through intermediate multivortex states with edge half antivortices. These nanostructured disks are promising to create multilevel memories and patterned arrays with multilevel high frequency permeability characteristic features. Moreover, the proposed nanostructuring approach allows for gaining effective metastable states in small ferromagnetic elements, especially given the crucial role of half antivortex states in the optimization of spin torque induced magnetization reversal in vortex nano oscillators and possibly other spintronic devices. Similar nanostructuring strategies could be implemented for dots of different geometries (triangles, squares) and thicknesses with multivortex metastable states.

Authors acknowledge E. Begun and R. Sachser for their help in preparing test samples. This work has been supported by the Spanish MINECO (MAT2012-32743) and Comunidad de Madrid (P2013/MIT2850). Computational capabilities of CCC-UAM (SVORTEX) and  FPI-UAM fellowship  (A. Lara) and also gratefully acknowledged.

REFERENCES

[1] O. Tchernyshyov, G-W. Chern, Phys. Rev. Lett., **95**, 197204 (2005).

[2] A. Pushp, T. Phung, C. Rettner, B.P. Hughes, H. Yang, L.Thomas, S.S.P. Parkin, Nature Physics, **9** 505 (2013).

[3] K-S. Lee, M-W. Yoo, Y-S. Choi, S-K. Kim, Phys. Rev. Lett., **106**, 147201 (2011).

[4] D. J. Clarke, O. A. Tretiakov, G.-W. Chern, Ya. B. Bazaliy, O. Tchernyshyov, Phys. Rev. B **78**, 134412 (2008).

[5] Nanomagnetism and Spintronics ($2^{nd}$ edition). Edited by Teruya Shinjo. Elsevier 2014.

[6] S. Kaka,  M. R. Pufall, W. H. Rippard, T. J. Silva, S. E. Russek, J. A. Katine, Nature, **437**, 389 (2005).

[7] A. A. Awad, G. R. Aranda, D. Dieleman, K.Y. Guslienko, G.N. Kakazei, B. A. Ivanov, F.G. Aliev, Appl. Phys. Lett., **97**, 132501 (2010).

[8] K.Y. Guslienko, J. Nanosci. Nanotechn., **8**, 2745 (2008).

[9] A. Hamadeh, N. Locatelli, V. V. Naletov, R. Lebrun, G. de Loubens, J. Grollier, O. Klein, and V. Cros,  Appl. Phys. Lett., **104**, 022408 (2014).

[10] G.E. Rowlands and I. Krivorotov, Phys. Rev. B **86**, 094425 (2012).

[11] A. Dussaux, E. Grimaldi, B. Rache Salles, A. S. Jenkins, A. V. Khvalkovskiy, P. Bortolotti, J. Grollier, H. Kubota, A. Fukushima, K. Yakushiji, S. Yuasa, V. Cros and A. Fert, Appl. Phys. Lett. **105**, 022404 (2014).

[12] J. Li, A. Tan, K. W. Moon, A. Doran, M. A. Marcus, A. T. Young, E. Arenholz[3], S. Ma[1], R. F. Yang, C. Hwang and Z. Q. Qiu, Appl. Phys. Lett. 104, 262409 (2014)

[13] J. A. J. Burgess, A. E. Fraser, F. Fani Sani, D. Vick, B.D. Hauer, J.P. Davis, M. R. Freeman, Science, **339** 1051 (2013).

[14] T. Uhlig, M. Rahm, C. Dietrich, R. Hollinger, M. Heumann, D. Weiss, and J. Zweck. Phys. Rev. Lett., **95**, 237205 (2005).

[15] G.M. Wysin, J. Phys.: Condens. Matter **22** 376002 (2010).

[16] M. Rahma, J. Biberger, V. Umansky, D. Weiss, JAP, **93,** 7429 (2003).

[17] V. P Kravchuk, D. D Sheka, F. G Mertens and Y. Gaididei, J. Phys. D: Appl. Phys. **44** 285001 (2011).

[18] M. Rahm, J. Stahl, W. Wegscheider, and D. Weiss, Appl. Phys. Lett., **85**, 1553 (2004).

[19] A. R. Pereira, A. R. Moura, W. A. Moura-Meloa, D. F. Carneiro, S. A. Leonel, and P. Z. Coura, JAP **101**, 034310 (2007).

[20] D. Toscano, S. A. Leonel, P. Z. Coura, F. Sato, R. A. Dias, B. V. Costa, Appl. Phys. Lett., **101**, 252402 (2012).

[21] F. Hoffmann, G. Woltersdorf, K. Perzlmaier, A. N. Slavin, V. S. Tiberkevich, A. Bischof, D. Weiss, and C. H. Back, Phys. Rev. **B76**, 014416 (2007).

[22] P. Vavassori, R. Bovolenta, V. Metlushko, B. Ilic, J. Appl. Phys. **99**, 053902 (2006).

[23] M. Rahm, J. Stahl, D. Weiss Appl. Phys. Lett., **87**, 182107 (2005).

[24] F.G. Aliev, A. Awad, D. Dieleman, A. Lara, V. Metlushko, and K.Y. Guslienko, Phys. Rev. **B84**, 144406 (2011)

[25] F.G. Aliev, D. Dieleman, A.A. Awad, A.Asenjo, O. Iglesias-Freire, M.García-Hernández, and V.Metlushko IEEE Explore, Electromagnetics in Advanced Applications (ICEAA), 160 (2010) DOI 10.1109/ICEAA.2010.5652137 Proceeding of 2010.

[26] N. Wang, X. L. Wang, and A. Ruotolo, IEEE TRANSACTIONS ON MAGNETICS, **47**. 1970 (2011).

[27] In the FEBID process the electron beam parameters were 5 kV / 1 nA and the process pressure was $1.2 \times 10^{-5}$ mbar. Before the deposition, the chamber was evacuated down to $7 \times 10^{-6}$ mbar. A visual inspection of the samples quality was done only several hours after the deposition process, in order to avoid autocatalytic dissociation of the residual Co on the substrate surface [33].

[28] F.G. Aliev, J. Sierra, A. Awad, G. Kakazei, D.-S. Han, S.-K. Kim, V. Metlushko, B. Ilic, and K.Y. Guslienko, Phys. Rev. **B79**, 174433 (2009).

[29] L. Serrano-Ramón, R. Córdoba, L.A. Rodríguez, C. Magén, E. Snoeck, C. Gatel, I. Serrano, M. R. Ibarra and J.M. De Teresa, ACS Nano **5**, 7781 (2011).

[30] M.J. Donahue and D.G. Porter. OOMMF User's Guide, Version 1.0. The parameters used in the micromagnetic simulations for Co are: 30 nm thickness, 2000 nm dot diameter, 5x5x30 $nm^3$ cell size in the $x$, $y$, and $z$ directions, correspondingly. The exchange stiffness is $1.4x10^{-11}$ J/m, the saturation magnetization $M_S$=$1.4x10^6$ A/m and the Gilbert damping constant $\alpha$=0.1. The equilibrium criterion for changing the field is that every simulation cell precesses slower than 1° per ns [35]. The current is assumed to be parallel to the electric field and independent of the magnetic field.

[31] See supplementary material at [URL to be added by AIP] for details.

[32] D. Herranz, A. Gomez-Ibarlucea, M. Schäfers, A. Lara, G. Reiss and F. G. Aliev, Appl. Phys. Lett. **99**, 062511 (2011).

[33] A. Chanthbouala, R. Matsumoto, J. Grollier, V. Cros, A. Anane, A. Fert, A. V. Khvalkovskiy, K.A. Zvezdin, K. Nishimura, Y. Nagamine, H. Maehara, K. Tsunekawa, A. Fukushima and S. Yuasa, Nature Physics **7**, 626 (2011)

[34] K. Muthukumar, H.O. Jeschke, R. Valenti, E. Begun, J. Schwenk, F. Porrati, M. Huth, Beilstein J. Nanotech. **3**, 546 (2012).

[35] D. Grujicic and B. Pesic, J. Magn. Magn. Mat. **285**, 303 (2005).

# FIGURE CAPTIONS

**Figure 1**

(a) The top left panel shows an optical image of the high frequency probe and the gold contacts to which it is attached. The other three panels show SEM images of the Co dots deposited by FEBID between the gold contacts. Three cases are shown: without, with two and with three nanoholes.

(b) Time evolution of the total energy of a hole-free dot after switching off the magnetic field (applied in the $x$ direction) that saturated it. Depending on the damping parameter $\alpha$ the system evolves differently. A high enough damping can stabilize metastable states. The different states, as deduced from the micromagnetic simulations, are indicated by arrows (in-plane magnetization, $M_x$ and $M_y$) and colors (vertical magnetization component, $M_z$).

**Figure 2**

(a) Measured AMR and frequency-averaged magnetic dynamic susceptibility (blue) for different nanohole configurations (see inset sketches) as functions of the applied magnetic field. Blue lines indicate magnetic transitions whose signatures appear in both types of measurements. Applied field and current directions are indicated by arrows.

(b) Simulated AMR curves of the respective dots. Arrows indicate the direction of variation of the applied field. Dashed blue arrows indicate the states shown in the first two columns of Fig. 3, with the color of the arrow head indicating which curve it points to, red or black. Only states with fundamentally different structures have been exemplified in Fig. 3.

The bottom axis shows fields normalized by the nucleation field of a double vortex (positive values correspond to nucleation going from positive to negative fields). The top axis (blue letters) represent the applied field (same values for all graphs in the same column).

**Figure 3**

Micromagnetic simulation results corresponding to the states labeled with blue numbers in Fig. 2. The black lines with arrows represent the magnetization direction, while the color scale represents the exchange energy density. The areas with high exchange energy density are DWs, V and HAVs. The right column shows a zoom of the different topological defects present in one of these states (bottom left), namely edge HAV at the dot edge, V, and several edge HAVs in a nanohole.

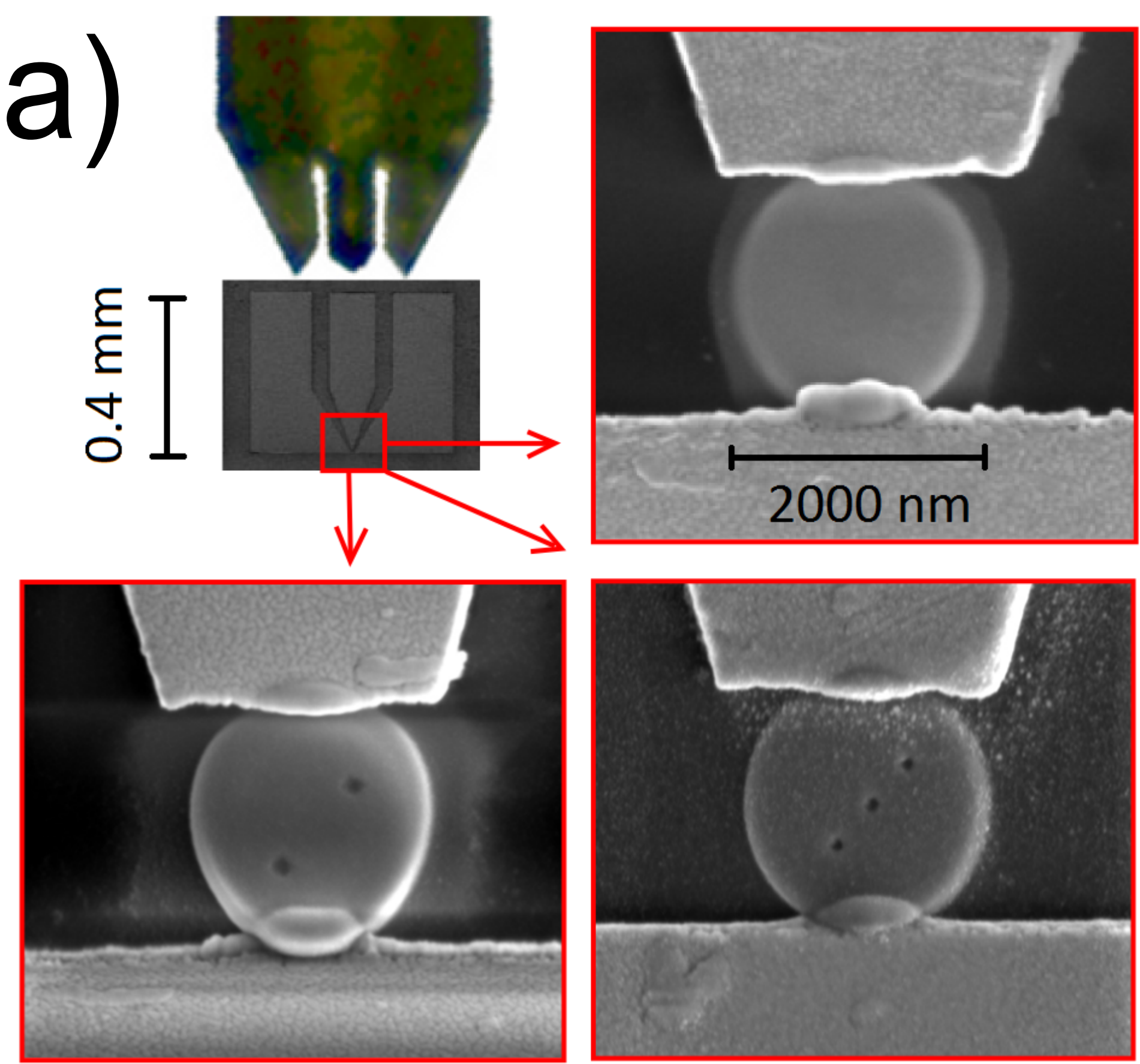

a)

0.4 mm

2000 nm

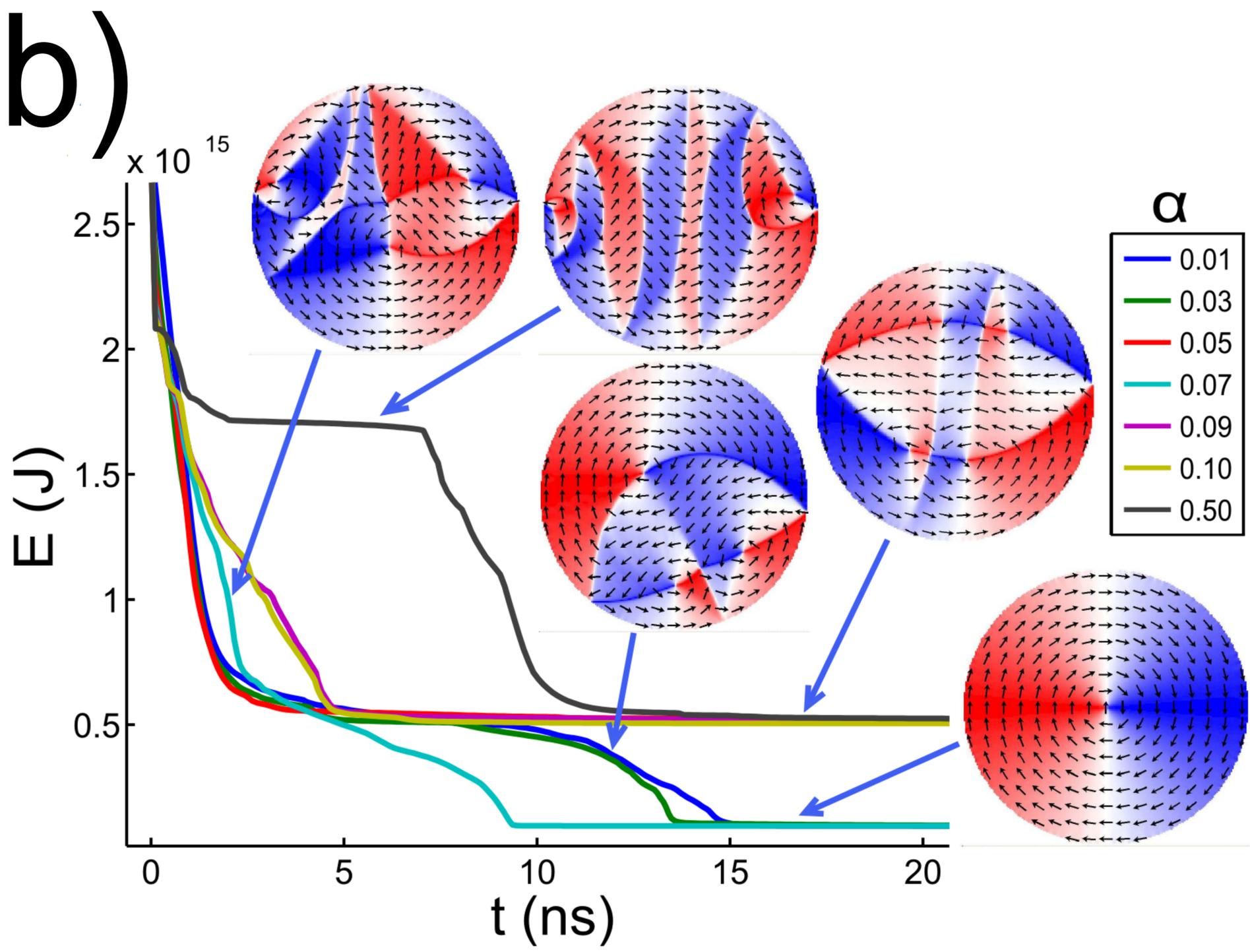

b)

$E$ (J)

$\times 10^{15}$

| α |
|---|
| 0.01 |
| 0.03 |
| 0.05 |
| 0.07 |
| 0.09 |
| 0.10 |
| 0.50 |

$t$ (ns)

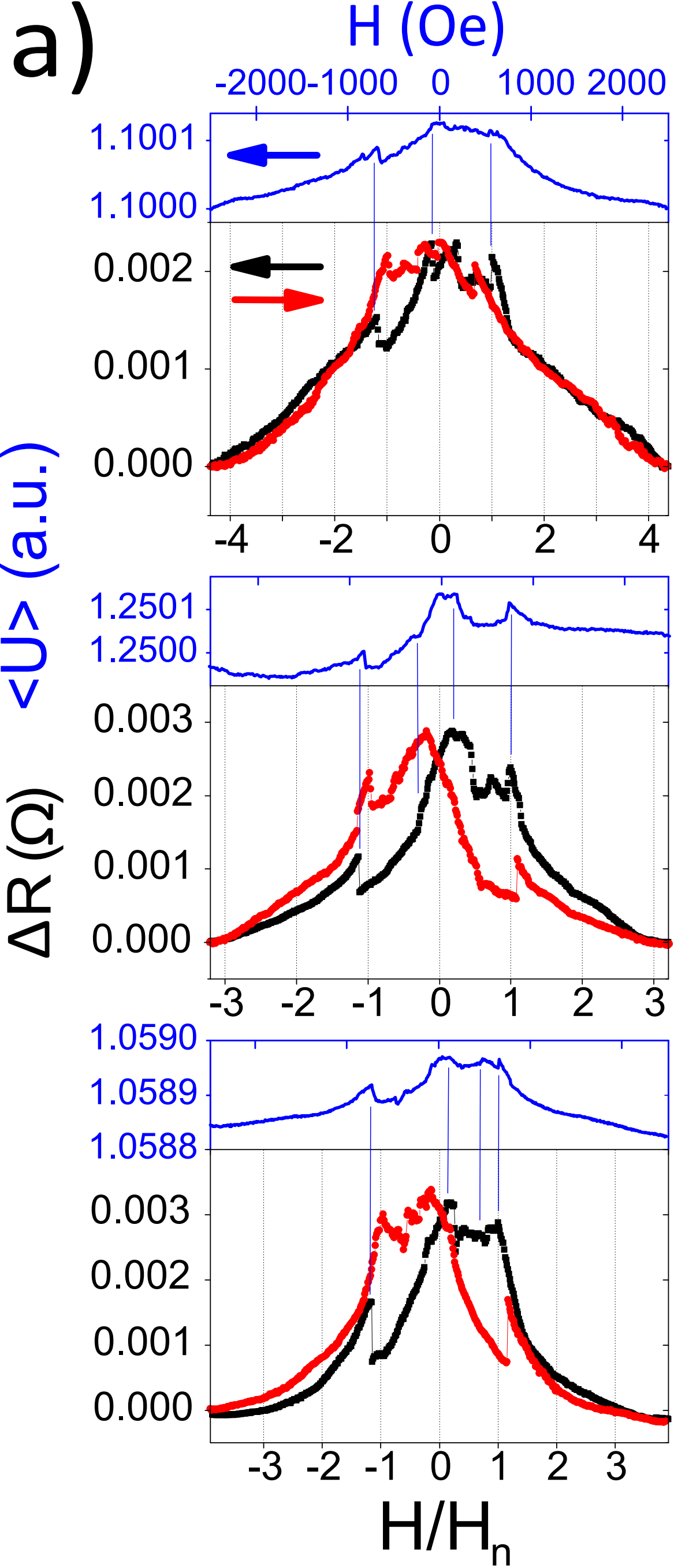
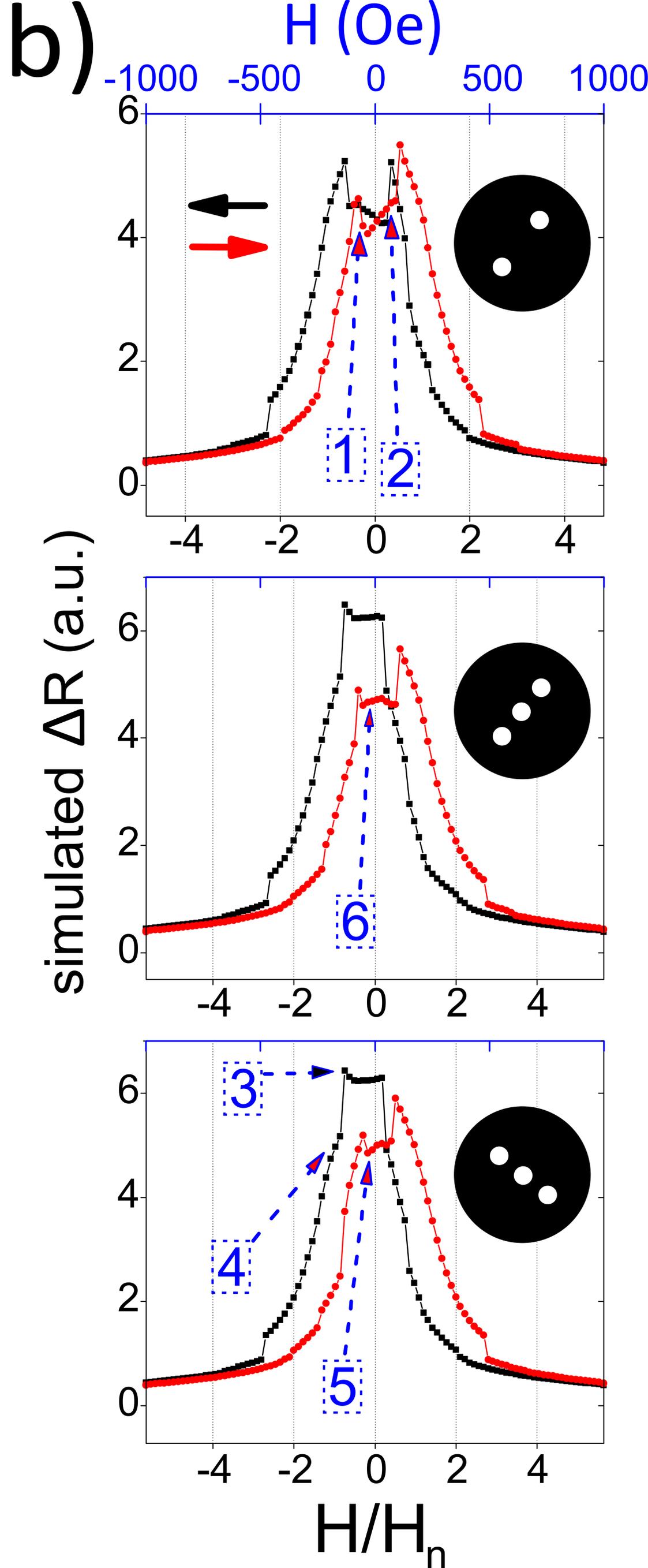

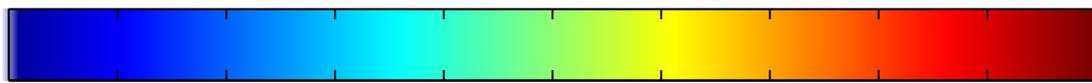

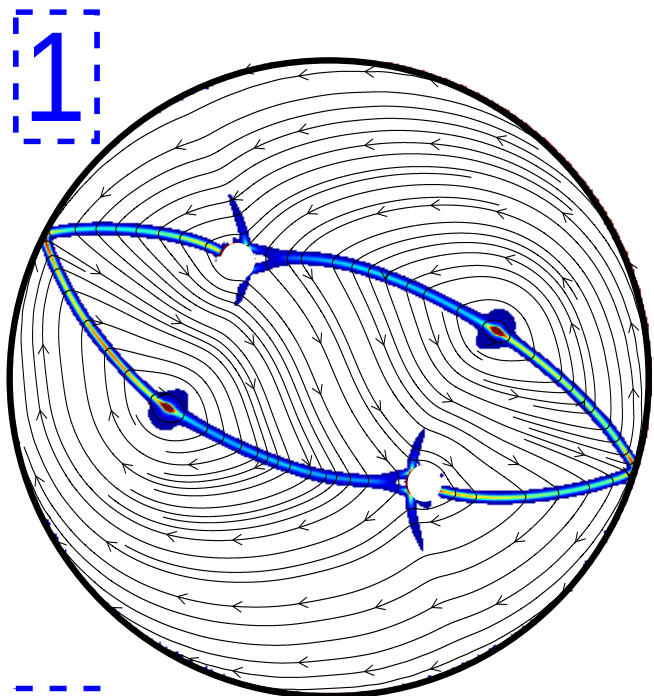
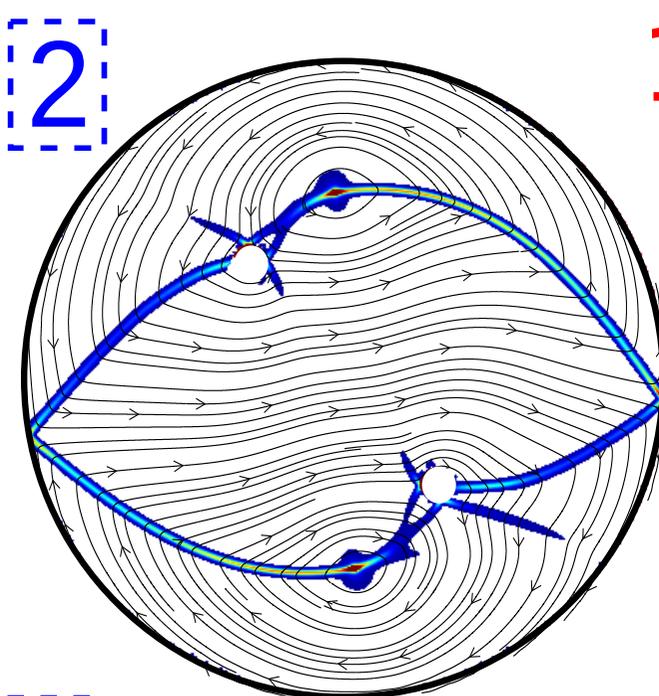
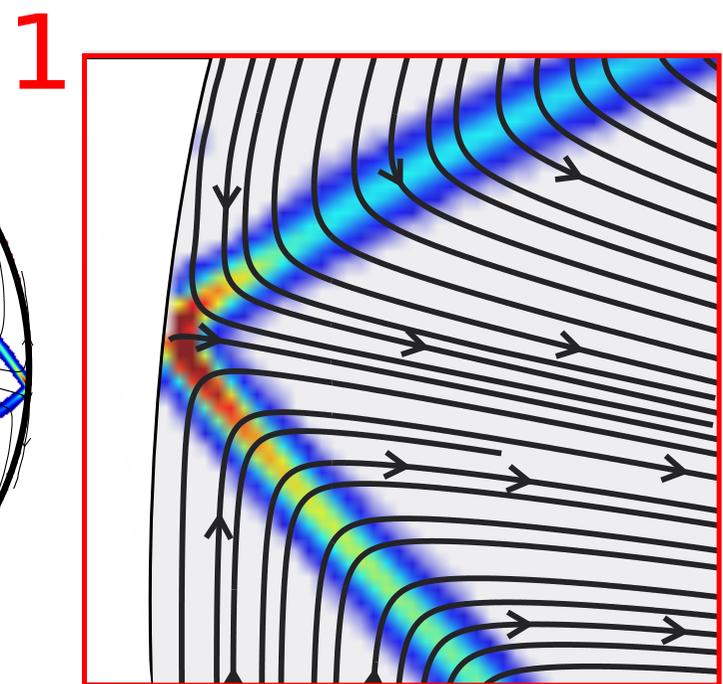

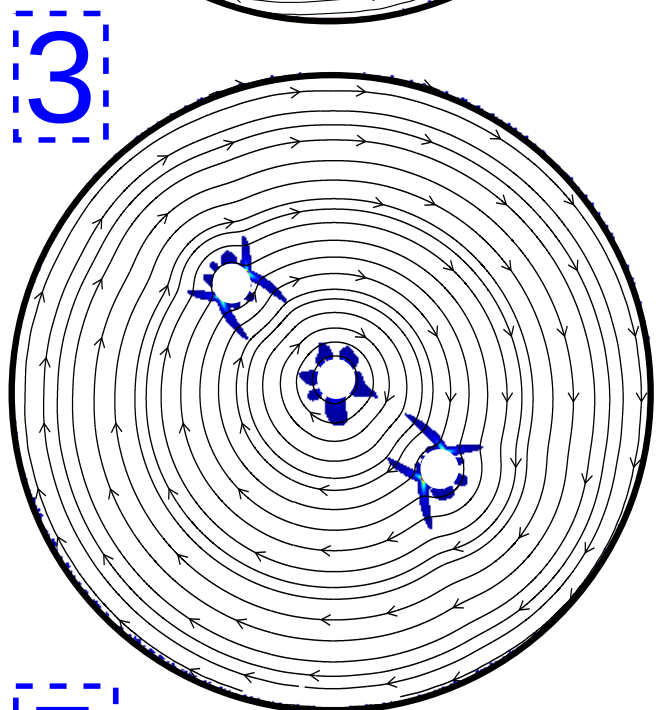
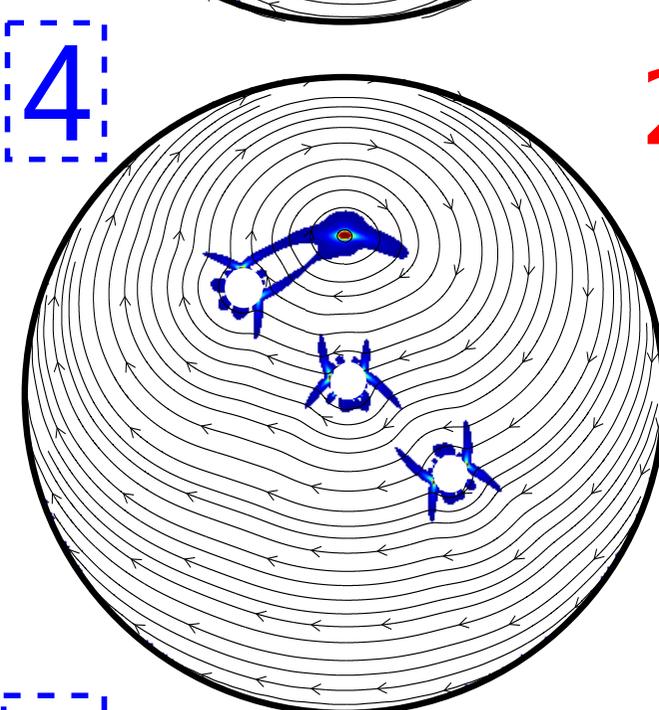
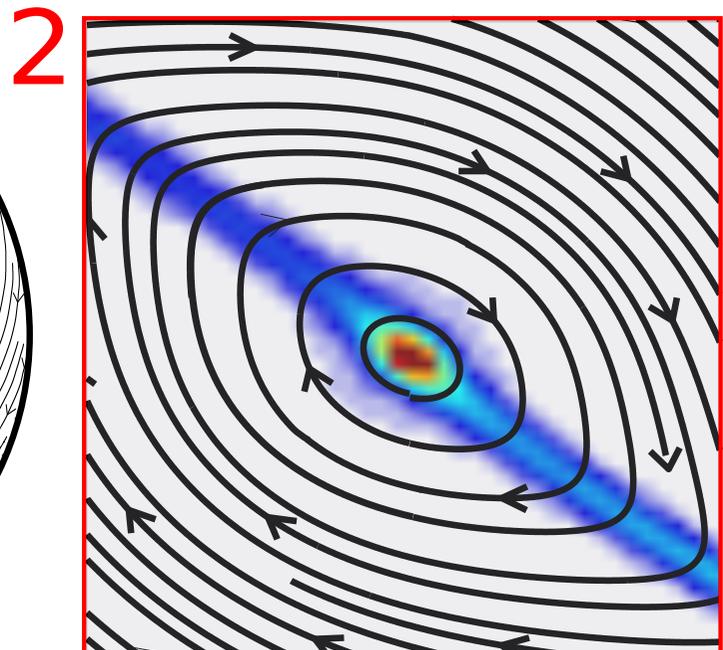

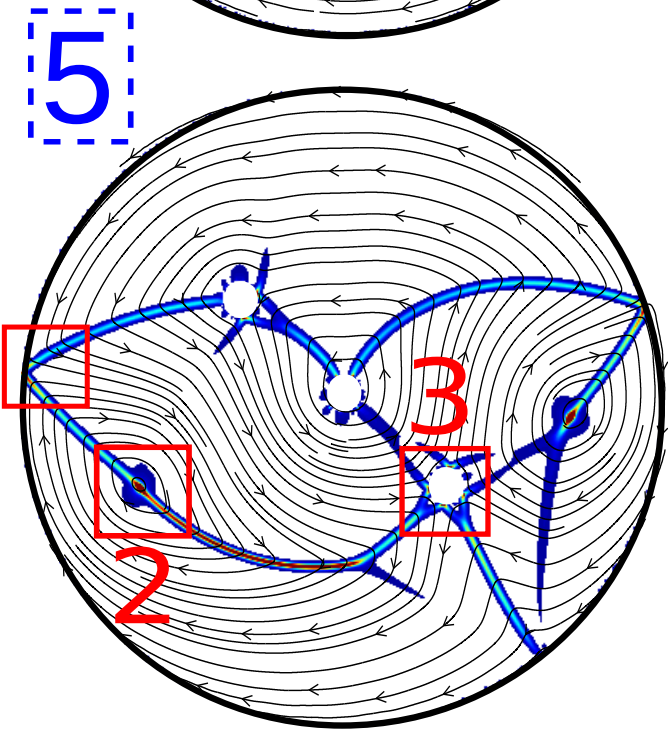
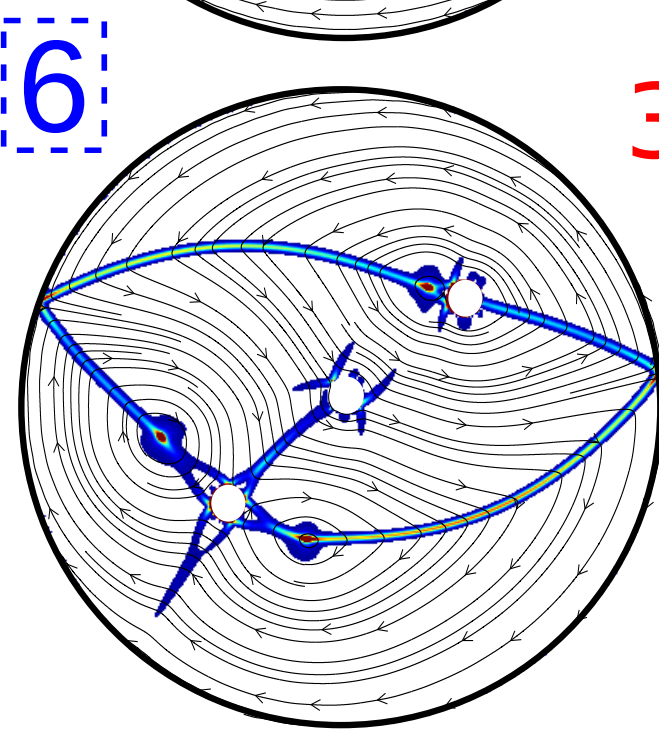
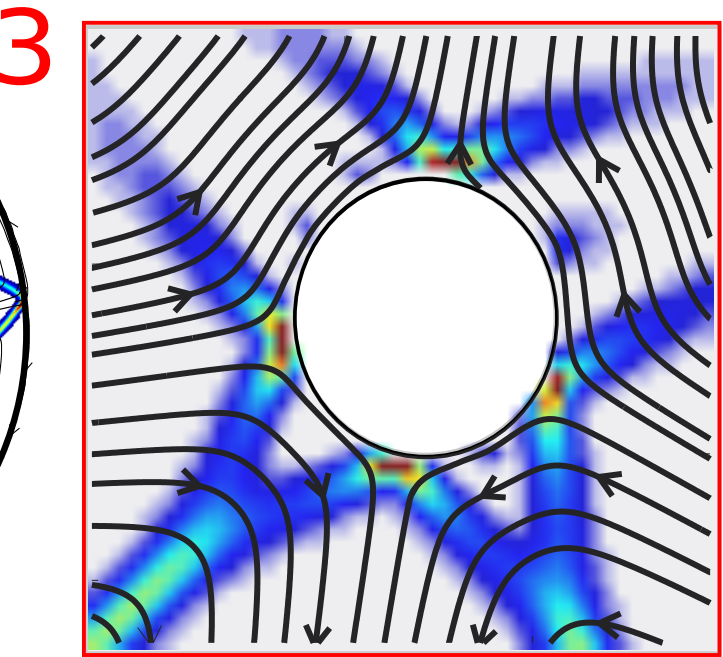